\documentclass{revtex4-2}%
\usepackage[latin9]{inputenc}
\usepackage{amsmath}
\usepackage{amssymb}
\usepackage{graphicx}
\usepackage[unicode=true,  bookmarks=false,  breaklinks=false,pdfborder={0 0
1},backref=section,colorlinks=false]{hyperref}
\usepackage{amsfonts}
\usepackage{subfigure}
\usepackage{cleveref}
\usepackage{listings}
\usepackage{xcolor}
\usepackage{array}
\usepackage{url}%
\hypersetup{
colorlinks,linkcolor=blue,citecolor=blue,urlcolor=blue}
\makeatletter
\@ifundefined{textcolor}{}
{
\definecolor{BLACK}{gray}{0}
\definecolor{WHITE}{gray}{1}
\definecolor{RED}{rgb}{1,0,0}
\definecolor{GREEN}{rgb}{0,1,0}
\definecolor{BLUE}{rgb}{0,0,1}
\definecolor{CYAN}{cmyk}{1,0,0,0}
\definecolor{MAGENTA}{cmyk}{0,1,0,0}
\definecolor{YELLOW}{cmyk}{0,0,1,0}
}

\makeatother
\begin{document}
\preprint{CTP-SCU/2021025}
\title{Testing Kerr Black Hole Mimickers with Quasi-Periodic Oscillations from GRO J1655-40}
\author{Xin Jiang}
\email{xjang@stu.scu.edu.cn}
\author{Peng Wang}
\email{pengw@scu.edu.cn}
\author{Houwen Wu}
\email{iverwu@scu.edu.cn}
\author{Haitang Yang}
\email{hyanga@scu.edu.cn}
\affiliation{Center for Theoretical Physics, College of Physics, Sichuan University,
Chengdu, 610064, China}

\begin{abstract}
The measurements of quasi-periodic oscillations (QPOs) provide a quite
powerful tool to test the nature of astrophysical black hole candidates in the
strong gravitational field regime. In this paper, we use QPOs within the
relativistic precession model to test a recently proposed family of rotating
black hole mimickers, which reduce to the Kerr metric in a limiting case, and
can represent traversable wormholes or regular black holes with one or two
horizons, depending on the values of the parameters. In particular, assuming
that the compact object of GRO J1655-40 is described by a rotating black hole
mimicker, we perform a $\chi$-square analysis to fit the parameters of the
mimicker with two sets of observed QPO frequencies from GRO J1655-40. Our
results indicate that although the metric around the compact object of GRO
J1655-40 is consistent with the Kerr metric, a regular black hole with one
horizon is favored by the observation data of GRO J1655-40.

\end{abstract}
\maketitle
\tableofcontents




\section{Introduction}

The first observations of gravitational waves by LIGO \cite{Abbott:2016blz}
and the first image of a black hole in the galaxy M87 \cite{Akiyama:2019cqa}
have ushered us into a new era of\ testing general relativity (GR) in the
strong gravity regime. On the other hand, quasi-periodic oscillations (QPOs)
are observed in the X-ray flux from black hole and neutron star X-ray binary
systems, and detected as narrow peaks in the power density spectrum
\cite{vanderKlis:2000ca}. QPOs are believed to be associated with motion and
accretion-related timescales in a region of order the Schwarzschild radius
around the compact object, which makes QPOs excellent probes of the strong
gravitational field regime
\cite{Berti:2015itd,Psaltis:2008bb,Psaltis:2009xf,Jusufi:2020odz,Azreg-Ainou:2020bfl}%
. In black hole systems, the observed frequencies of QPOs range from mHz to
hundreds of Hz. While low frequency QPOs $\left(  \lesssim30\text{
}\mathrm{Hz}\right)  $ are commonly observed from black hole X-ray binaries
\cite{Ingram:2019mna}, high frequency QPOs $\left(  \gtrsim60\text{
}\mathrm{Hz}\right)  $ are very rare. In fact, the Rossi X-ray Timing Explorer
(RXTE), operational between 1996 and 2012, first detected high frequency QPOs
in black hole systems \cite{Morgan:1997su,remillard1999rxte}. Interestingly, a
pair of simultaneous high-frequency QPOs was first discovered in the X-ray
flux from GRO J1655-40 by RXTE \cite{strohmayer2001discovery}. It was noted
that the frequencies of the two high-frequency QPOs are in a $3$:$2$ ratio,
suggesting a resonance between orbital and epicyclic motion of accreting
matter near the innermost stable circular orbit (ISCO) of black holes
\cite{Abramowicz:2001bi}. Later, three simultaneous QPO frequencies,
consisting of two higher frequencies and one lower frequency, were also
observed from the X-ray data of GRO J1655-40 \cite{Motta:2013wga}.

On the theoretical side, various models have been proposed to explain QPOs by
relating them to the orbital and epicyclic frequencies of geodesics, such as
the relativistic precession model (RPM) \cite{Stella:1999sj}, the tidal
disruption model \cite{Cadez:2008iv}, the parametric resonance model
\cite{Rebusco:2004ba}, the resonance model
\cite{Abramowicz:2001bi,Kluzniak:2002bb,Abramowicz:2004je}, the warped disk
oscillation model \cite{Kato:2007ny} and the non-axisymmetric disk oscillation
model \cite{Bursa:2004qk,Torok:2010rk}. With these models, the observation
data of QPO frequencies have been used to constrain the parameters of X-ray
binary systems and test the nature of various gravity theories
\cite{Beer:2001cg,Johannsen:2010bi,Kulkarni:2011cy,Pappas:2012nt,Bambi:2012pa,Bambi:2013fea,Maselli:2014fca,Suvorov:2015yfv,Bambi:2016iip,Chen:2016yey,Allahyari:2021bsq,Chen:2021jgj,Banerjee:2021aln,DeFalco:2021btn,Rink:2021mwt}%
.

Particularly, it showed that the X-ray data of GRO J1655-40, especially the
QPO triplet, fit nicely in the RPM, and the mass and spin of the compact
object of GRO J1655-40 can be precisely determined \cite{Motta:2013wga}.
Remarkably, the inferred mass is in great agreement with the dynamical mass
measurement \cite{Beer:2001cg}. The RPM was originally proposed to explain
QPOs in low-mass X-ray binaries with a neutron star \cite{Stella:1997tc}, and
later extended to systems with stellar-mass BH candidates \cite{Stella:1999sj}%
. In the RPM, QPO frequencies are assumed to be related to fundamental
frequencies of a test particle orbiting a central object. The twin higher
frequencies are regarded as the azimuthal frequency $\nu_{\phi}$ and the
periastron precession frequency $\nu_{\text{per}}$ of quasi-circular orbits in
the innermost disk region, respectively. The low-frequency QPO is identified
as the nodal precession frequency $\nu_{\text{nod}}$, which is emitted at the
same radius where the twin higher frequencies are generated.

On the other hand, curvature singularities can be formed during a
gravitational collapse. It is commonly believed that singularities can be
avoided through quantum gravitational effects. Consequently, since Bardeen
proposed the first regular black hole \cite{Bardeen:1968non}, constructing and
studying classical black holes without singularities have been a topic of
considerable interest in GR and astrophysics communities due to their
non-singular property
\cite{Roman:1983zza,Hayward:2005gi,Bardeen:2014uaa,Frolov:2016pav,Cano:2018aod,Bardeen:2018frm,Carballo-Rubio:2018pmi,Carballo-Rubio:2018jzw}%
. Recently, Simpson and Visser proposed a static and spherically symmetric
regular spacetime described by the line element (dubbed the SV metric
henceforth),
\begin{equation}
ds^{2}=-\left(  1-\frac{2M}{\sqrt{r^{2}+\ell^{2}}}\right)  dt^{2}+\left(
1-\frac{2M}{\sqrt{r^{2}+\ell^{2}}}\right)  ^{-1}dr^{2}+\left(  r^{2}+\ell
^{2}\right)  \left(  d\theta^{2}+\sin^{2}\theta d\phi^{2}\right)  ~,
\label{eq:SV}%
\end{equation}
where $M\geq0$ represents the ADM mass, and $\ell>0$ is a parameter
responsible for regularizing the center singularity
\cite{Simpson:2018tsi,Simpson:2019cer,Simpson:2019oft}. The most appealing
feature of the SV metric $\left(  \ref{eq:SV}\right)  $ is that it can
smoothly interpolate between a regular black hole for $\ell<2M$ and a wormhole
for $\ell\geq2M$. In the limit of $\ell=0$, the SV metric reduces to the
ordinary Schwarzschild spacetime. Subsequently, the properties of the SV
metric were discussed, e.g., quasinormal modes \cite{Churilova:2019cyt},
precessing and periodic geodesic motions \cite{Zhou:2020zys}, gravitational
lensing \cite{Nascimento:2020ime,Cheng:2021hoc,Tsukamoto:2021caq} and shadows
\cite{Bronnikov:2021liv,Guerrero:2021ues}. To make the SV metric more relevant
to realistic situations, the SV metric $\left(  \ref{eq:SV}\right)  $ was
generalized to a family of rotating black hole mimickers, dubbed the rotating
SV metric, which may represent a rotating wormhole and a rotating regular
black hole with one or two horizons \cite{Mazza:2021rgq}. The strong
gravitational lensing and the shadows of the rotating SV metric have been
investigated in \cite{Islam:2021ful} and \cite{Shaikh:2021yux}, respectively.

In this paper, we test gravity with QPOs frequencies observed from GRO
J1655-40 within the RPM for the rotating SV metric. The content of this paper
is as follows. After we briefly review the RPM and the rotating SV metric in
sections \ref{sec:EF} and \ref{sec:RSVM}, respectively, the epicyclic
frequencies of the rotating SV are computed in section \ref{sec:RSVM}. In
section \ref{sec:CRSVMQPO}, we use the data of GRO J1655-40 to put constraints
on the parameters of the rotating SV metric. Section \ref{sec:CON} is devoted
to our conclusions. Throughout the paper, we use units in which $G=c=1$.

\section{Epicyclic Frequencies}

\label{sec:EF}

In this section, we consider the timelike geodesic equations in a stationary
and axially symmetric spacetime, and then derive the expressions of the
epicyclic frequencies. The metric of a stationary and axially symmetric
spacetime which satisfies the circularity condition is given by
\cite{Chandrasekhar:1985kt}
\begin{equation}
ds^{2}=g_{tt}dt^{2}+g_{rr}dr^{2}+2g_{t\phi}dtd\phi+g_{\theta\theta}d\theta
^{2}+g_{\phi\phi}d\phi^{2}, \label{eq:metric}%
\end{equation}
where the metric $g_{\mu\nu}$ is a function of $r$ and $\theta$, and we drop
the coordinate dependence of the metric functions to simplify the notation.
For a massive particle travelling along a time-like world line $x^{\mu}\left(
\tau\right)  $ with $\tau$ being the proper time, the four-velocity $U^{\mu}$
is defined by $U^{\mu}=\dot{x}^{\mu}=dx^{\mu}/d\tau$, which satisfies $U^{\mu
}U_{\mu}=-1$. Due to the stationarity and axisymmetry, the metric $\left(
\ref{eq:metric}\right)  $ admits two Killing vectors,%
\begin{equation}
K^{\mu}=(\partial_{t})^{\mu}=(1,0,0,0)\text{ and }R^{\mu}=(\partial_{\phi
})^{\mu}=(0,0,0,1).
\end{equation}
The two Killing vectors correspond to two conserved quantities of geodesic
motion,%
\begin{align}
E  &  =-K_{\mu}\frac{dx^{\mu}}{d\tau}=-g_{tt}\dot{t}-g_{t\phi}\dot{\phi
},\nonumber\\
L_{z}  &  =R_{\mu}\frac{dx^{\mu}}{d\tau}=g_{t\phi}\dot{t}+g_{\phi\phi}%
\dot{\phi},
\end{align}
which can be interpreted as the energy per unit mass and the angular momentum
per unit mass along the axis of symmetry, respectively. In terms of $E$ and
$L_{z}$, one can express $\dot{t}$ and $\dot{\phi}$ as%
\begin{equation}
\dot{t}=\frac{g_{\phi\phi}E+g_{t\phi}L_{z}}{g_{t\phi}^{2}-g_{tt}g_{\phi\phi}%
}\text{, }\dot{\phi}=\frac{g_{t\phi}E+g_{tt}L_{z}}{g_{tt}g_{\phi\phi}%
-g_{t\phi}^{2}}\text{.}%
\end{equation}
Using the above equations, we can rewrite $U^{\mu}U_{\mu}=-1$ as
\begin{equation}
g_{rr}\dot{r}^{2}+g_{\theta\theta}\dot{\theta}^{2}=V_{\text{eff}}\left(
r,\theta\right)  , \label{eq:gVeff}%
\end{equation}
where the effective potential $V_{\text{eff}}\left(  r,\theta\right)  $ is
defined as%
\begin{equation}
V_{\text{eff}}(r,\theta)=\frac{E^{2}g_{\phi\phi}+2EL_{z}g_{t\phi}+L_{z}%
^{2}g_{tt}}{g_{t\phi}^{2}-g_{tt}g_{\phi\phi}}-1.
\end{equation}

We consider a circular geodesic at $r=\bar{r}$ on the equatorial plane with
$\theta=\pi/2$, which means that the effective potential may develop a double
root at $r=\bar{r}$ on the equatorial plane, i.e., $V_{\text{eff}}(\bar{r}%
,\pi/2)=\partial_{r}V_{\text{eff}}(\bar{r},\pi/2)=0$. Along the circular
orbit, the angular velocity of the particle measured by an observer at
infinity is defined by%
\begin{equation}
\Omega_{\phi}=\frac{d\phi}{dt}=\frac{\dot{\phi}}{\dot{t}}=\left.
-\frac{g_{t\phi}E+g_{tt}L_{z}}{g_{\phi\phi}E+g_{t\phi}L_{z}}\right\vert
_{r=\bar{r},\theta=\pi/2}\text{.}\label{eq:av}%
\end{equation}
Solving $V_{\text{eff}}(\bar{r},\pi/2)=0$ with eqn. $\left(  \ref{eq:av}%
\right)  $ gives the specific energy and angular momentum of the particle,
\begin{align}
E &  =\left.  -\frac{g_{tt}+g_{t\phi}\Omega_{\phi}}{\sqrt{-g_{tt}-2g_{t\phi
}\Omega_{\phi}-g_{\phi\phi}\Omega_{\phi}^{2}}}\right\vert _{r=\bar{r}%
,\theta=\pi/2},\nonumber\\
L_{z} &  =\left.  \frac{g_{t\phi}+g_{\phi\phi}\Omega_{\phi}}{\sqrt
{-g_{tt}-2g_{t\phi}\Omega_{\phi}-g_{\phi\phi}\Omega_{\phi}^{2}}}\right\vert
_{r=\bar{r},\theta=\pi/2}.
\end{align}
Solving $\partial_{r}V_{\text{eff}}(\bar{r},\pi/2)=0$ for $\Omega_{\phi}$, one
obtains the angular velocity of the particle,%
\begin{equation}
\Omega_{\phi}=\left.  \frac{-\partial_{r}g_{t\phi}\pm\sqrt{\left(
\partial_{r}g_{t\phi}\right)  ^{2}-\left(  \partial_{r}g_{tt}\right)  \left(
\partial_{r}g_{\phi\phi}\right)  }}{\partial_{r}g_{\phi\phi}}\right\vert
_{r=\bar{r},\theta=\pi/2},
\end{equation}
where the sign $+/-$ corresponds to a prograde/retrograde orbit. The stability
of the circular orbit is determined by the sign of $\partial_{r}%
^{2}V_{\text{eff}}(\bar{r},\pi/2)$, i.e., $\partial_{r}^{2}V_{\text{eff}}%
(\bar{r},\pi/2)>0$ $\Leftrightarrow$ unstable and $\partial_{r}^{2}%
V_{\text{eff}}(\bar{r},\pi/2)<0$ $\Leftrightarrow$ stable. The transition
between stable and unstable circular orbits, which is determined by
$\partial_{r}^{2}V_{\text{eff}}(r_{\text{ISCO}},\pi/2)=0$, is the ISCO, which
is located at $r=r_{\text{ISCO}}$ on the equatorial plane.

To derive the epicyclic frequencies associated with the circular orbit, we
consider small perturbations of the orbit in both the radial and the vertical
directions,%
\begin{equation}
r(t)=\bar{r}+\delta r(t)\text{,}\quad\theta(t)=\frac{\pi}{2}+\delta\theta(t).
\label{eq:pert}%
\end{equation}
Inserting eqn. $\left(  \ref{eq:pert}\right)  $ into eqn. $\left(
\ref{eq:gVeff}\right)  $ yields the differential equations for the
perturbations $\delta r(t)$ and $\delta\theta(t)$,%
\begin{equation}
\frac{d^{2}\delta r(t)}{dt^{2}}+\Omega_{r}^{2}\delta r(t)=0\text{,}\quad
\frac{d^{2}\delta\theta(t)}{dt^{2}}+\Omega_{\theta}^{2}\delta\theta(t)=0,
\end{equation}
where the frequencies of the oscillations are
\begin{equation}
\Omega_{r}^{2}=\left.  -\frac{1}{2g_{rr}\dot{t}^{2}}\frac{\partial
^{2}V_{\text{eff}}\left(  r,\theta\right)  }{\partial r^{2}}\right\vert
_{r=\bar{r},\theta=\pi/2},\text{ }\Omega_{\theta}^{2}=\left.  -\frac
{1}{2g_{\theta\theta}\dot{t}^{2}}\frac{\partial^{2}V_{\text{eff}}\left(
r,\theta\right)  }{\partial\theta^{2}}\right\vert _{r=\bar{r},\theta=\pi/2}.
\label{eq:efre}%
\end{equation}
We then define $\nu_{\phi}=\Omega_{\phi}/2\pi$, $\nu_{r}=\Omega_{r}/2\pi$ and
$\nu_{\theta}=\Omega_{\theta}/2\pi$ as the azimuthal, radial and vertical
epicyclic frequencies, respectively. The periastron precession frequency
$\nu_{\text{per}}$ and the nodal precession frequency $\nu_{\text{nod}}$ are
defined by $\nu_{\text{per }}=\nu_{\phi}-\nu_{r}$ and $\nu_{\text{nod }}%
=\nu_{\phi}-\nu_{\theta}$, respectively.

\section{Rotating Simpson-Visser Metric}

\label{sec:RSVM}

In \cite{Mazza:2021rgq}, a rotating generalization of the static and
spherically symmetric metric $\left(  \ref{eq:SV}\right)  $ has been
constructed by employing the Newman--Janis procedure \cite{Newman:1965tw}.
This stationary and axially symmetric metric can describe a rotating
traversable wormhole and a rotating regular black hole with one or two
horizons. In particular, the rotating SV metric reads%
\begin{equation}
ds^{2}=-\left(  1-\frac{2M\sqrt{r^{2}+\ell^{2}}}{\Sigma}\right)  dt^{2}%
+\frac{\Sigma}{\Delta}dr^{2}+\Sigma d\theta^{2}-\frac{4Ma\sin^{2}\theta
\sqrt{r^{2}+\ell^{2}}}{\Sigma}dtd\phi+\frac{A\sin^{2}\theta}{\Sigma}d\phi
^{2}~, \label{eq:RSV}%
\end{equation}
with
\begin{align}
\Sigma &  =r^{2}+\ell^{2}+a^{2}\cos^{2}\theta,\nonumber\\
\Delta &  =r^{2}+\ell^{2}+a^{2}-2M\sqrt{r^{2}+\ell^{2}},\\
A  &  =\left(  r^{2}+\ell^{2}+a^{2}\right)  ^{2}-\Delta a^{2}\sin^{2}%
\theta,\nonumber
\end{align}
where $a$ is the spin parameter. The rotating SV metric will reduce to the SV
metric $\left(  \ref{eq:SV}\right)  $ if $a=0$ and to the Kerr metric if
$\ell=0$. Interestingly, the rotating SV metric is everywhere regular when
$\ell>0$ \cite{Mazza:2021rgq}.

The horizons of the rotating SV metric are determined by $\Delta=0$, whose
solutions are%
\begin{equation}
r_{\pm}=\sqrt{\left(  M\pm\sqrt{M^{2}-a^{2}}\right)  ^{2}-l^{2}}.
\end{equation}
As shown in \cite{Mazza:2021rgq}, the phases of the rotating SV metric are
determined by the existence of $r_{\pm}$. Specifically, the rotating SV metric represents

\begin{itemize}
\item a traversable wormhole: $M<a$ or $l>M+\sqrt{M^{2}-a^{2}}$;

\item a regular black hole with one horizon (RBH-I): $M-\sqrt{M^{2}-a^{2}%
}<l<M+\sqrt{M^{2}-a^{2}}$\ and $M>a$;

\item a regular black hole with two horizons (RBH-II): $l<M-\sqrt{M^{2}-a^{2}%
}$\ and $M>a$;

\item there limiting cases: a one-way wormhole with a null throat when
$l=M+\sqrt{M^{2}-a^{2}}$, a regular black hole with one horizon and a null
throat when $l=M-\sqrt{M^{2}-a^{2}}$ and an extremal regular black hole when
$M=a$.
\end{itemize}

\begin{figure}[tb]
\begin{minipage}[t]{0.3\linewidth}
		\centering
		\includegraphics[width=2.1in]{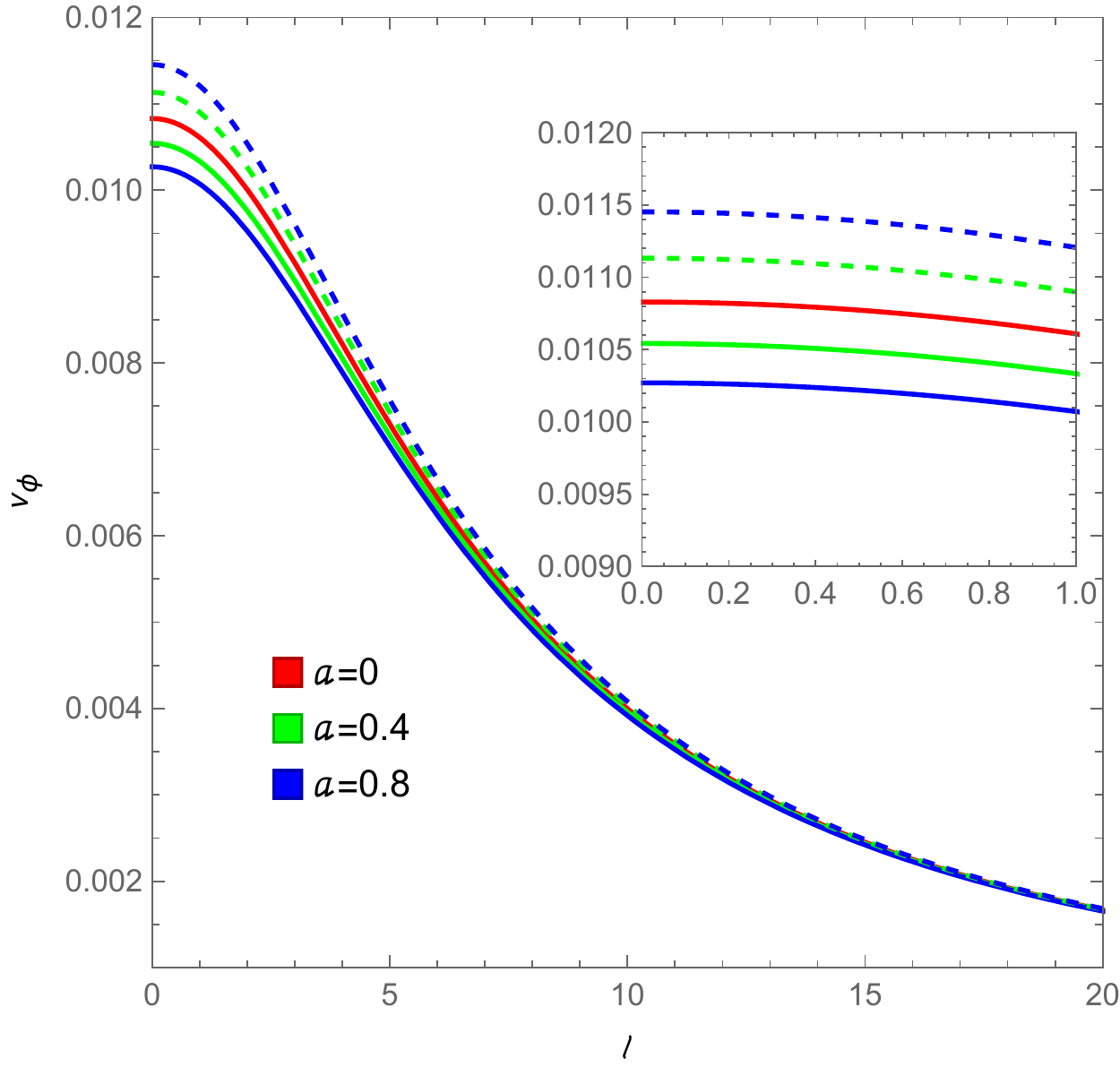}
	\end{minipage}
\begin{minipage}[t]{0.3\linewidth}
		\centering
		\includegraphics[width=2.1in]{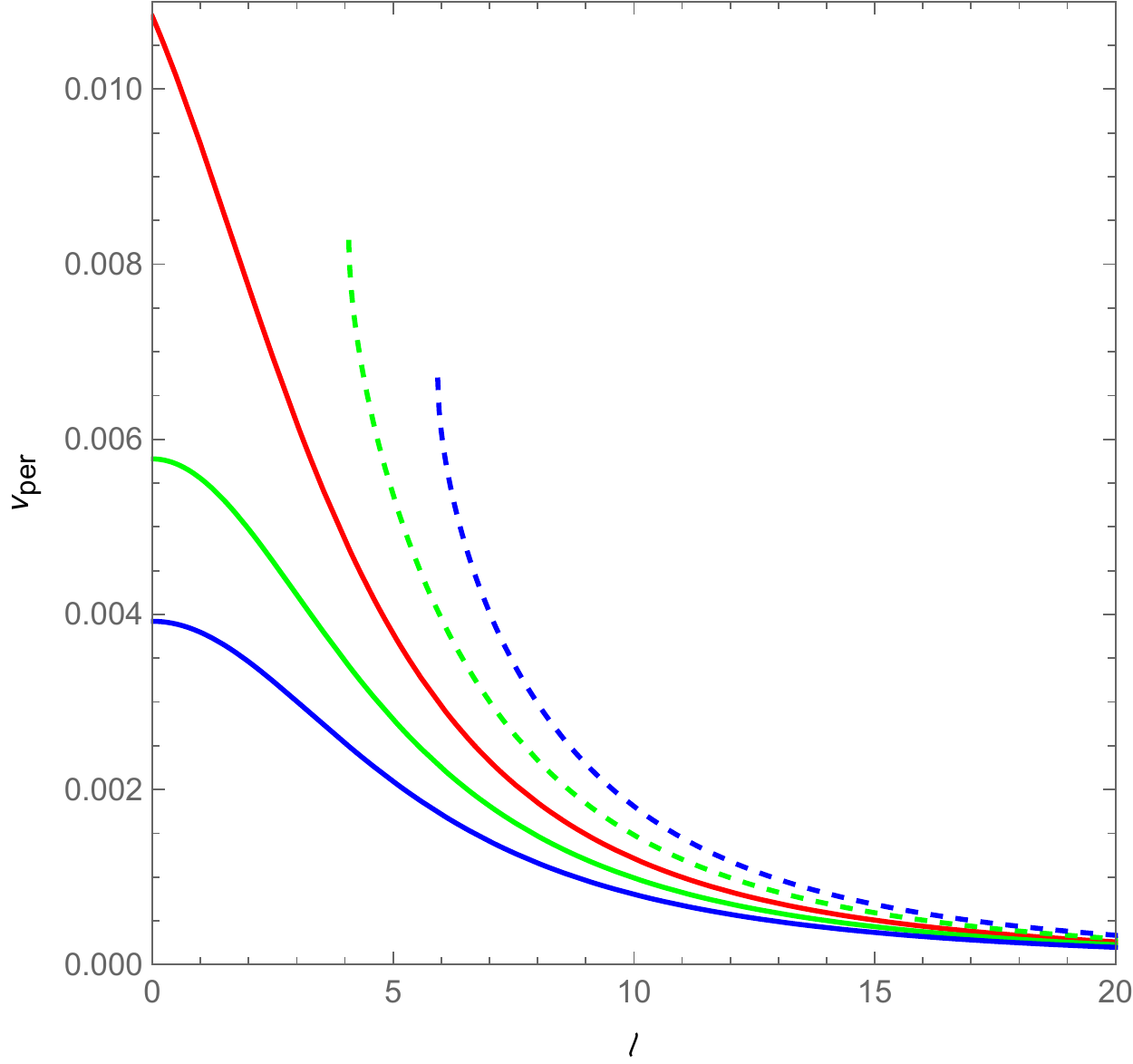}
	\end{minipage}
\begin{minipage}[t]{0.3\linewidth}
		\centering
		\includegraphics[width=2.1in]{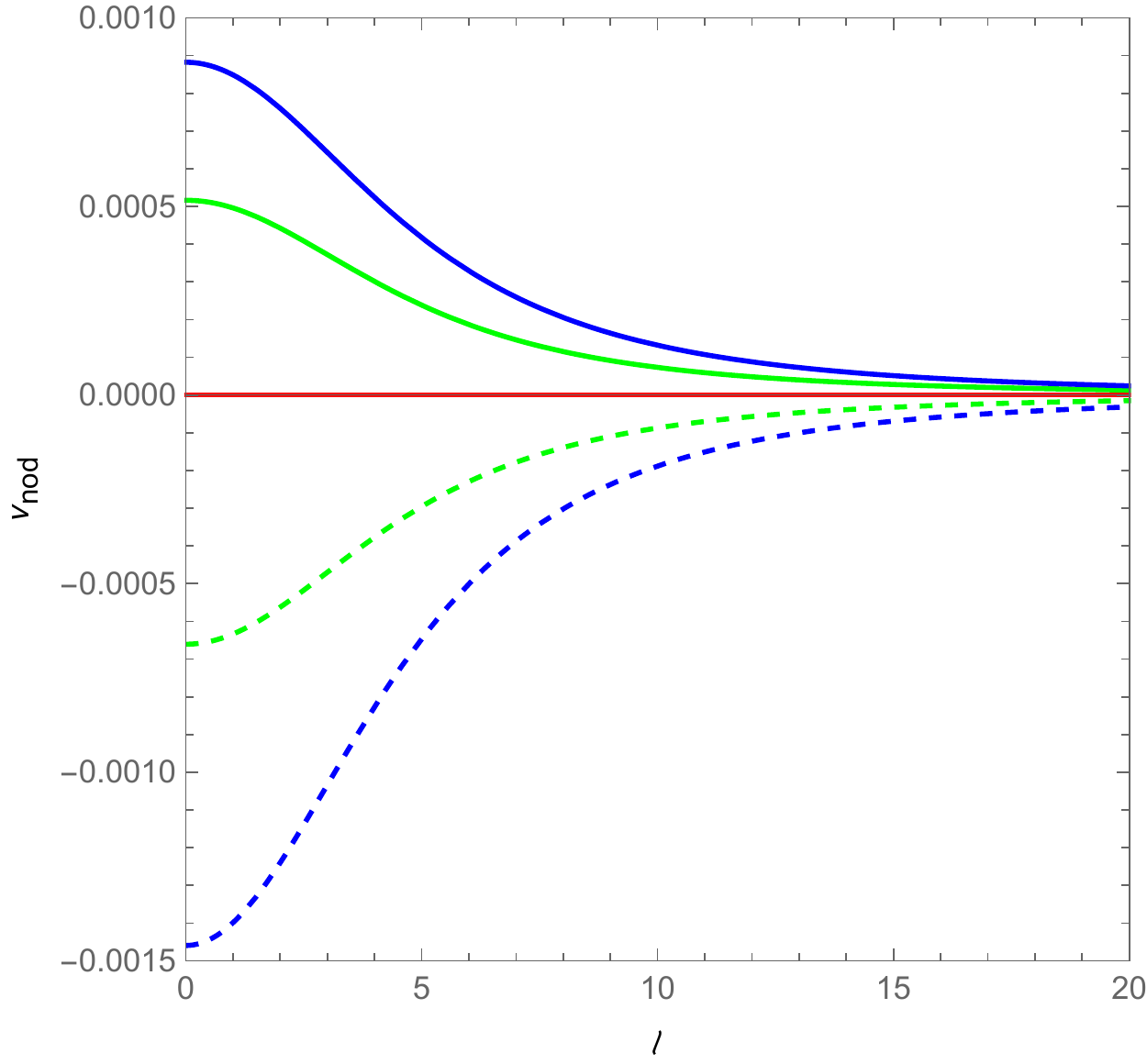}
	\end{minipage}
\caption{Plots of the azimuthal epicyclic frequency $\nu_{\phi}$ (left panel),
the periastron precession frequency $\nu_{\text{per}}$ (middle panel) and the
nodal precession frequency $\nu_{\text{nod}}$ (right panel) of prograde (solid
lines) and retrograde (dashed lines) orbits at $\bar{r}=6$ as a function of
$\ell$ for $a=0$ (red lines), $a=0.4$ (green lines) and $a=0.8$ (blue lines)
in the rotating SV metric with $M=1$. The $a=0$ case corresponds to the SV
metric, which has $\nu_{\text{nod}}=0$ due to spherical symmetry. Except
$\nu_{\text{nod}}$ with $a=0$, the magnitudes of $\nu_{\phi}$, $\nu
_{\text{per}}$ and $\nu_{\text{nod}}$ decrease as $\ell$ grows with a fixed
$a$. For the retrograde orbits in the rotating SV metric with $a>0$, the
values of $\nu_{\text{nod}}$ are shown to be negative.}%
\label{fig:QPOs}%
\end{figure}

For a circular orbit at $r=\bar{r}$ on the equatorial plane, substituting the
rotating SV metric $\left(  \ref{eq:RSV}\right)  $ into eqns. $\left(
\ref{eq:av}\right)  $ and $\left(  \ref{eq:efre}\right)  $ gives epicyclic
frequencies in the rotating SV metric,%
\begin{align}
\nu_{\phi}  &  =\frac{1}{2\pi}\frac{M^{1/2}}{(\bar{r}^{2}+\ell^{2})^{3/4}\pm
aM^{1/2}},\nonumber\\
\nu_{r}  &  =\frac{r\nu_{\phi}}{(\bar{r}^{2}+\ell^{2})^{1/2}}\sqrt{1-\frac
{6M}{(\bar{r}^{2}+\ell^{2})^{1/2}}-\frac{3a^{2}}{(\bar{r}^{2}+\ell^{2})}%
\pm\frac{8aM^{1/2}}{(\bar{r}^{2}+\ell^{2})^{3/4}}},\\
\nu_{\theta}  &  =\nu_{\phi}\sqrt{1\mp\frac{4aM^{1/2}}{(\bar{r}^{2}+\ell
^{2})^{3/4}}+\frac{3a^{2}}{(\bar{r}^{2}+\ell^{2})}},\nonumber
\end{align}
which can be used to constrain the four parameters $\ell$, $r$, $M$, and $a$
of the rotating SV metric. Here, the top/bottom row of the $\pm$ and $\mp$
signs corresponds to the orbit co-rotating/counter-rotating with the
spacetime. To illustrate the dependence of the epicyclic frequencies on $\ell
$, we plot $\nu_{\phi}$, $\nu_{\text{per}}$ and $\nu_{\text{nod}}$ as a
function of $\ell$ for various values of $a$ in FIG. \ref{fig:QPOs}, where
$M=1$ and $\bar{r}=6$. The prograde and retrograde cases are represented by
solid and dashed lines, respectively. The left panel shows that, for a fixed
$a$, the azimuthal epicyclic frequency $\nu_{\phi}$ decreases with $\ell$
increasing in both prograde and retrograde cases, whereas $\nu_{\phi}$ of the
prograde orbit is smaller than that of the retrograde orbit. The periastron
precession frequency $\nu_{\text{per}}$ is displayed in the middle panel, and
also decreases as $\ell$ increases for both prograde and retrograde orbits.
Like $\nu_{\phi}$, the retrograde orbits have larger $\nu_{\text{per}}$. Note
that retrograde circular orbits of radius $\bar{r}=6$ do not exist when $\ell$
is small enough. However, as shown in the right panel, the nodal precession
frequency $\nu_{\text{nod}}$ of the prograde/retrograde orbits
decreases/increases as $\ell$ increases for a given $a$. More interestingly,
when $a>0$, $\nu_{\text{nod}}$ of the retrograde orbits is negative while that
of the prograde orbits is positive. Finally, it is noteworthy that the ISCO
radius $r_{\text{ISCO}}$ is determined by \cite{Mazza:2021rgq},
\begin{equation}
r_{\text{ISCO}}^{2}+\ell^{2}-6M\sqrt{r_{\text{ISCO}}^{2}+\ell^{2}}\pm
8a\sqrt{M\sqrt{r_{\text{ISCO}}^{2}+\ell^{2}}}=3a^{2},
\end{equation}
where $+/-$ is associated with the prograde/retrograde ISCO.

\section{Constraining Rotating Simpson-Visser Metric by Quasi-Periodic
Oscillations}

\label{sec:CRSVMQPO}

In this section, we use the RPM along with the QPO frequencies from GRO
J1655-40 to put constraints on the parameters of the rotating SV metric. GRO
J1655-40 is an X-ray binary, consisting of a primary star and a compact
companion \cite{Orosz:1996cg}. The measurement of the X-ray spectrum was found
to exhibit type-C low-frequency QPOs and simultaneous high-frequency QPOs,
which are observed in pairs and therefore dubbed lower and upper
high-frequency QPOs \cite{strohmayer2001discovery,Motta:2013wga}. In
particular, we consider two sets of QPOs with the observed frequencies based
on the RXTE observations \cite{Motta:2013wga},
\begin{align}
\nu_{1\mathrm{U}}=441\mathrm{~Hz},\quad &  \sigma_{1\mathrm{U}}=2\mathrm{~Hz}%
,\nonumber\\
\nu_{1\mathrm{L}}=298\mathrm{~Hz},\quad &  \sigma_{1\mathrm{L}}=4\mathrm{~Hz}%
,\nonumber\\
\nu_{1\mathrm{C}}=17.3\mathrm{~Hz},\quad &  \sigma_{1\mathrm{C}}%
=0.1\mathrm{~Hz}%
\end{align}
and
\begin{align}
\nu_{2\mathrm{U}}=451\mathrm{~Hz},\quad &  \sigma_{2\mathrm{U}}=5\mathrm{~Hz}%
,\nonumber\\
\nu_{2\mathrm{L}}=  &  ~-~,\quad\nonumber\\
\nu_{2\mathrm{C}}=18.3\mathrm{~Hz},\quad &  \sigma_{2\mathrm{C}}%
=0.1\mathrm{~Hz}~.
\end{align}

\begin{figure}[ptb]
\begin{minipage}[t]{0.3\linewidth}
		\centering
		\includegraphics[width=2.1in]{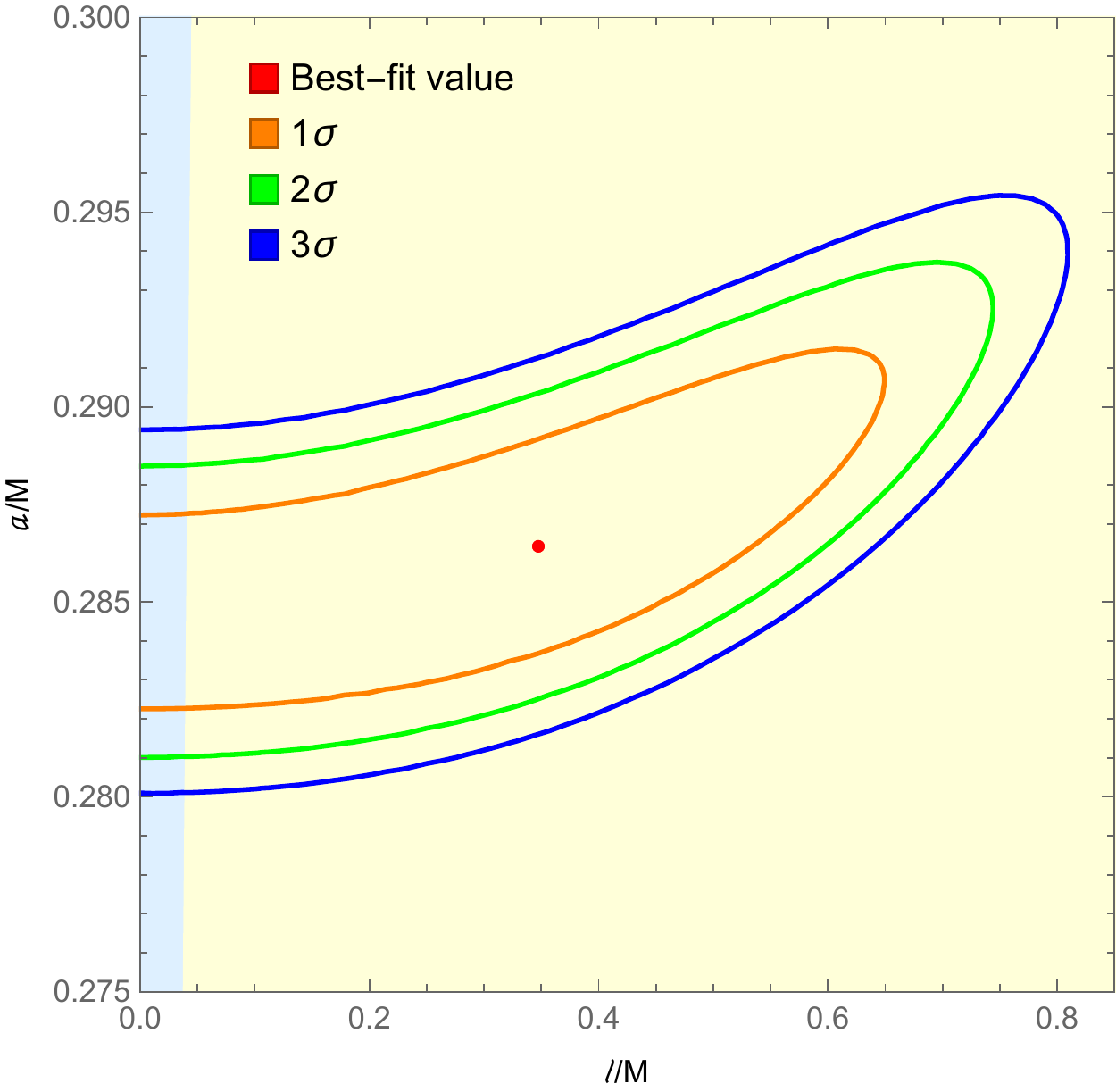}
	\end{minipage}
\begin{minipage}[t]{0.3\linewidth}
		\centering
		\includegraphics[width=2.1in]{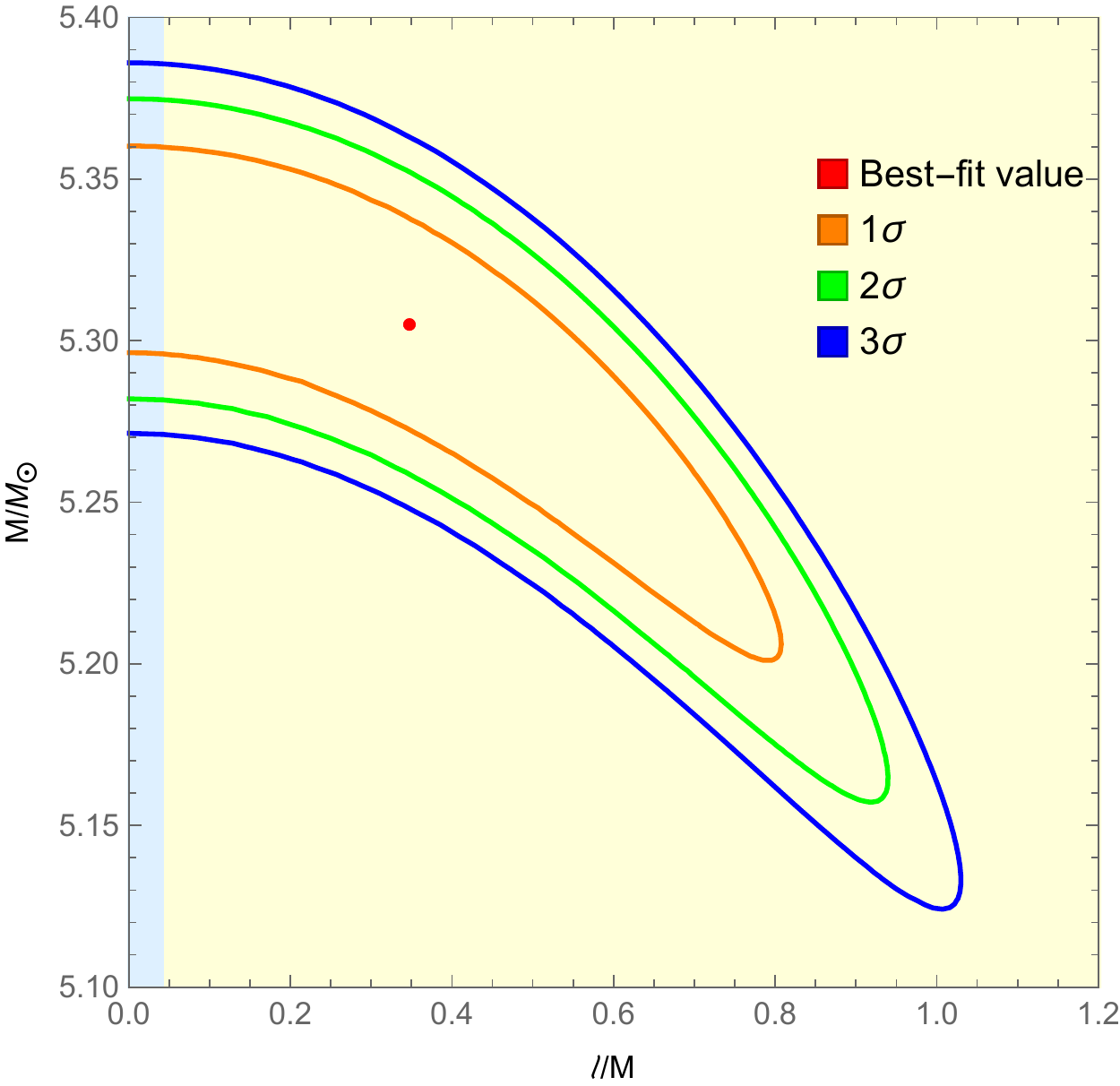}
	\end{minipage}
\begin{minipage}[t]{0.3\linewidth}
		\centering
		\includegraphics[width=2.1in]{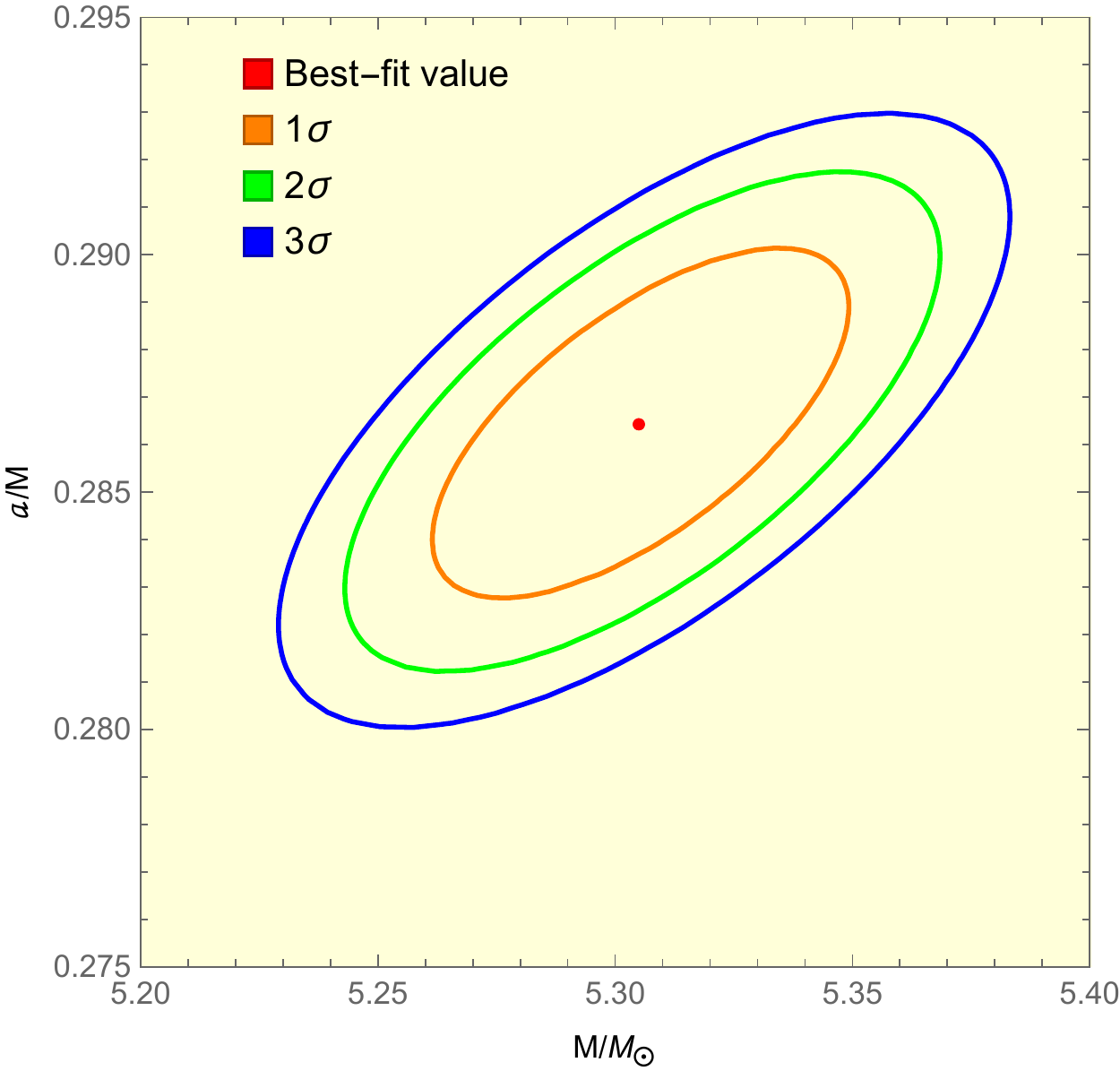}
	\end{minipage}
\caption{Estimate of the mass $M/M_{\odot}$, of the spin parameter $a/M$ and
of the parameter $\ell/M$ of the rotating SV metric, which describes the
spacetime around the compact object in GRO J1655-40, from interpreting the
observations of QPOs in the RPM. Specifically, the best-fit values (red dots)
and the contour levels of $1\sigma$ (orange lines), $2\sigma$ (green lines)
and $3\sigma$ (blue lines) for the parameters are shown in the $\ell/M$-$a/M$
(left panel), $\ell/M$-$M/M_{\odot}$ (middle panel) and $M/M_{\odot}$-$a/M$
(right panel) planes. In the yellow and blue regions, the rotating SV metric
represents regular black hole solutions with one and two horizons,
respectively. The observed frequencies of the QPOs in GRO J1655-40 are
consistent with a Kerr black hole (the rotating SV metric with $\ell=0$), but
they also allow for large deviations from the Kerr black hole solution. In
particular, a regular black hole with one horizon is favored over the other
phases of the rotating SV metric, e.g., a regular black hole with two horizons
and a traversable wormhole.}%
\label{fig:estimate}%
\end{figure}

In the RPM, three simultaneous QPO frequencies are generated at the same
radial coordinate in the accretion disk. The upper high-frequency QPOs
correspond to the azimuthal epicyclic frequency $\nu_{\phi}$, the lower
high-frequency QPOs to the periastron precession frequency $\nu_{\text{per}}$,
and the low-frequency QPOs to the nodal precession frequency $\nu_{\text{nod}%
}$. Moreover, it is reasonable to assume that the above two sets of QPOs
result from two circular orbits of different radii, i.e., $r_{1}$ and $r_{2}$.
In short, we have five free parameters: the mass $M$, the spin parameter $a$,
the $\ell$ parameter, and the radii $r_{1}$ and $r_{2}$ corresponding to the
QPOs with three frequencies and two frequencies, respectively. To obtain the
estimate of the five parameters from the observed QPO frequencies, we follow
the procedure used in \cite{Bambi:2013fea,Allahyari:2021bsq,Chen:2021jgj} and
perform a $\chi$-square analysis with%
\begin{equation}
\chi^{2}(M,a,\ell,r_{1},r_{2})=\frac{(\nu_{1\phi}-\nu_{1\mathrm{U}})^{2}%
}{\sigma_{1\mathrm{U}}}+\frac{(\nu_{1\text{per}}-\nu_{1\mathrm{L}})^{2}%
}{\sigma_{1\mathrm{L}}}+\frac{(\nu_{1\text{nod}}-\nu_{1\mathrm{C}})^{2}%
}{\sigma_{1\mathrm{C}}}+\frac{(\nu_{2\phi}-\nu_{2\mathrm{U}})^{2}}%
{\sigma_{2\mathrm{U}}}+\frac{(\nu_{2\text{nod}}-\nu_{2\mathrm{C}})^{2}}%
{\sigma_{2\mathrm{C}}}~,
\end{equation}
the minimum of which, $\chi_{\text{min}}^{2}$, occurs at the best estimate of
$M$, $a$ $\ell$, $r_{1}$ and $r_{2}$. The range of the parameters at a
confidence level ($\mathrm{C.L.}$) is determined by the interval
$\chi_{\text{min}}^{2}+\Delta\chi^{2}$. In the case of five degrees of
freedom, the intervals with $\Delta\chi^{2}=5.89$, $11.29$ and $17.96$
correspond to $68.3\%$, $95.4\%$ and $99.7\%$ $\mathrm{C.L.}$, respectively,
which are the probability intervals designated as $1$, $2$, and $3$ standard
deviation limits, respectively.

Computing $\chi^{2}$, we find $\chi_{\text{min}}^{2}=$ $0.195$ and obtain the
best fits of the parameters of the rotating SV metric within $68.3\%$
credibility,%
\begin{equation}
M/M_{\odot}=5.305_{-0.028}^{+0.041}\text{, }a/M=0.286_{-0.002}^{+0.003}\text{,
}\ell/M=0.347_{-0.347}^{+0.011}. \label{eq:bestfit}%
\end{equation}
Note that our result is consistent with the measurement of the mass $M$ by
optical and infrared observations, which give $M=5.4\pm0.3M_{\odot}$
\cite{Beer:2001cg}. We present the best estimate and the $1\sigma$, $2\sigma$
and $3\sigma$ contour levels of $M/M_{\odot}$, $a/M$ and $\ell/M$ in FIG.
\ref{fig:estimate}, where the yellow/blue regions represent the RBH-I/RBH-II
phases of the rotating SV metric. As shown in the left and middle panels,
while the hypothesis that the compact object of GRO J1655-40 is described by a
Kerr black hole is consistent with the interpretation of the QPOs' data in the
RPM, significant deviations from the Kerr metric are allowed. In fact, the
best-fit values are in the parametric region of the RBH-I phase, and the
regions within 1-, 2- and 3-standard deviation limits are almost in the RBH-I
region. Therefore, the observation of GRO J1655-40 favors a regular black hole
with one horizon if the compact object of GRO J1655-40 is described by the
rotating SV metric. The best-fit values of the radii of the circular orbits
associated with the two sets of QPOs are found to be $r_{1}%
=5.669M=1.1304r_{\text{ISCO}}$ and $r_{2}=5.563M=1.1094r_{\text{ISCO}}$,
respectively, where $r_{\text{ISCO}}=5.105M$ is the innermost stable circular
orbit evaluated for the rotating SV metric with the best-fit values $\left(
\ref{eq:bestfit}\right)  $. Consequently, the two circular orbits responsible
for generating the two sets of QPOs lie in the close vicinity of the ISCO, and
hence are in the strong-field region of the rotating SV metric.

\section{Conclusions}

\label{sec:CON}

In this paper, we explored potential deviations from the GR predictions of
astrophysical black holes using QPOs observed in the power density spectrum of
GRO J1655-40. Specially, we modelled the spacetime around the compact object
of GRO J1655-40 by the rotating SV metric, and interpreted the observed QPOs
within the RPM, which relates the QPO frequencies to epicyclic frequencies of
geodesics. The rotating SV metric reduces to a Kerr black hole in the limit of
$\ell=0$, and possesses multiple phases, e.g., a regular black hole with one
or two horizons and a traversable wormhole. To test the nature of the the
compact object of GRO J1655-40, we performed a $\chi^{2}$ analysis by fitting
the QPO frequencies computed in the RPM with the observations of two sets of
QPOs from GRO J1655-40. Our results show a preference towards a regular black
hole with one horizon compared to the Kerr black hole predicted in GR.

\begin{acknowledgments}
We thank Guangzhou Guo for his helpful discussions and suggestions. This work
is supported in part by NSFC (Grant No. 11875196, 11375121, 11947225 and 11005016).
\end{acknowledgments}

\bibliographystyle{unsrturl}
\bibliography{ref}

\end{document}